\documentclass{rrparticle}
\usepackage{graphicx}

\newcommand{\miktex}{\hbox{Mik\kern-.15em\TeX}}

\title{Constrained Path Monte Carlo method for spin $ 1/2 $ fermions at unitarity limit} 
\author[1]{F. Etminan}
\author[1,a]{M. M. Firoozabadi}
\affil[1]{Department of Physics,\\ Faculty of Sciences, University of Birjand,\\ Birjand 97175-615, Iran\\Email:{\em fetminan@birjand.ac.ir} Email:$^a$ {\em mfiroozabadi@birjand.ac.ir}}
 

\begin{document}
\maketitle
\begin{abstract}
We present calculations for spin $ 1/2 $ fermions at unitarity limit, where the effective range of the interaction is zero and the scattering length is infinite. We compute the ground-state energy for a system of 6, 10,14,18 and 20 particles, with equal numbers of up and down spins in a periodic cube in the full ground-state constrained-path Monte Carlo (CPMC) method using the extended, attractive Hubbard model. 
Our results in a careful extrapolation to the thermodynamic limit may suggest that the ratio of the ground-state energy to that of a free Fermi gas is $ \xi = 0.43(4) $, which can compare with recent experimental results and consistent with the fixed node Green's function Monte Carlo and novel lattice approaches results. We also obtain results for interactions with different effective ranges and find that the energy is consistent with a universal linear dependence on the product of the Fermi momentum and the effective range.
\end{abstract}

\section{\label{sec:intro} Introduction}
A model for strongly interacting fermions, which has absorbed a vast field of research \cite{hao2015}
is Fermi gas at unitary limit, zero-range attractive interaction and infinite scattering length.
The model is of interest in both condensed-matter and nuclear physics. The unitary limit in nuclear physics is realized when the interparticle spacing is about $ 5 - 10  $ fm, roughly $ 0.5 - 5 $\verb+%+  of normal nuclear matter density \cite{leeprb2006}, which is relevant to the physics of the inner crust of neutron stars. Therefore the unitary limit is relevant to the properties of cold dilute neutron matter \cite{leeprb2006,leeprb2007}. The reciprocation among experiment, theory, and computation has led to rapid advances \cite{ku2012,carlson2011}. An example is the evolution \cite{endres2013} of the calculation of the so-called Bertsch parameter, $ \xi $, at unitarity. The scaling properties in the unitarity limit are the same as those of a noninteracting Fermi gas. At zero temperature in the unitary limit there are no dimensional parameters other than the particle density. 

Let $ E_{N\uparrow,N\downarrow}^{0,free} $ be the ground-state energy for $ N_{\uparrow} $ up-spin and $ N_{\downarrow} $  down-spin free noninteracting fermions with equal masses in a periodic cube. We write $ E_{N\uparrow,N\downarrow}^{0} $  for the ground-state energy at unitarity for the same particle numbers, $ N_{\uparrow} $  and $ N_{\downarrow} $, and the same periodic cube volume. We then define energy ratio \cite{bour2011} 
\begin{equation}
\xi_{N_{\uparrow},N_{\downarrow}}=\frac{E_{N_{\uparrow},N_{\downarrow}}^{0}}{E_{N_{\uparrow},N_{\downarrow}}^{0,free}} 
\label{eq:en_ratio}.
\end{equation}

Quantitative comparisons is important for our understanding of many-body physics and provided a motivation for developments of both experimental and theoretical techniques. Many experiments and calculations have been performed for the unitary Fermi gas. Initial significant agreement achieved between calculation \cite{carlson2011}
and experiment \cite{ku2012,partridge2006} illustrates the tremendous progress towards understanding of strongly correlated quantum matter.

%which is more compatible with recent experimental results
More precise recent experiments have yielded $ 0.51(4) $ \cite{kinast2005}, $ 0.46(5) $, 0.435(15) \cite{joseph2007}, 0.41(2) \cite{luo2009}, and $ 0.41(1) $ \cite{navon2010}.

Fixed-node diffusion Monte Carlo (DMC) calculations \cite{forbes2011,gandolfi2011}
have always included a Bardeen- Cooper-Schrieffer \cite{bardeen1957}
(BCS) trial wave function to guide the Monte Carlo walk and provide the fixed node constraint \cite{anderson1976} needed to overcome the fermion sign problem. As is well known, these calculations provide an upper bound, with the current best value $ \xi=0.383(1) $ \cite{forbes2011,gandolfi2011}.

Several lattice simulations of two-component fermions in the unitarity limit by extrapolating the average energy at nonzero temperature to the zero-temperature limit is done. In which $ \xi $ is calculated between $ 0.07 $ and $ 0.42 $ \cite{lee2006b}. In Ref.~\cite{bulgac2006} a value for $ \xi $ is produced in the $ 0.3 $ to $ 0.5 $ interval. More recent lattice calculations extrapolated to zero temperature yield values of $ \xi=0.292(24) $ \cite{abe2009a} and $ \xi=0.37(5) $ \cite{bulgac2008}. In Refs.~ \cite{leeprb2006,lee2008} the values of $  \xi_{N,N} $ was estimated $ \xi=0.25(3) $ for $ N=3,5,7,9,11 $ fermions at lattice volumes $ 4^{3},5^{3},6^{3} $ in units of lattice spacing. Auxiliary-field Monte Carlo methods and Euclidean time projection is used to calculate ground-state energy of systems. 
In Ref. \cite{leeprc2008} this lattice calculation was improved using bounded continuous auxiliary fields. The results obtained were $ \xi_{55}=0.292(12) $ and $ \xi_{77}=0.329(5) $. Another non ab initio method called the symmetric heavy-light ansatz gives an estimate of $ \xi=0.31(1) $ in the continuum and thermodynamic limits \cite{lee2008}. Density-functional theory method which include shell effects suggests $ \xi=0.322(2) $ \cite{forbes2011}. Novel lattice approach for studying large numbers of fermions using different projection, sampling and measuring correlation functions produce a value of $ \xi_{N,N}=0.412(4) $ \cite{endres2013}. The most predicted values for $ \xi $, range from $ 0.3 $ to $ 0.4 $.

\section{ \label{sec:AFQMCM} Auxiliary Field Quantum Monte Carlo Methods}
Often the only suitable tool for microscopic calculations of strongly interacting many-body systems are Quantum Monte Carlo (QMC) methods.

For systems where there is a sign problem \cite{loh1990}, constraining the random walks in sampling the space of auxiliary fields, will led to considerable progress, these methods are called constrained-path Monte Carlo (CPMC) \cite{zhang1997,hao2013}. Sign problem arises from the combination of Pauli principle and the use of random sampling. If the system size or inverse temperature increased, signal-to-noise ratio will vanishing exponentially. The idea is to constrain the sign or phase of the overlap of the sampled Slater determinants with a trial wave function \cite{hao2013}. Applications to a variety of systems have shown that the methods are very accurate, even with simple trial wave functions taken directly from mean-field calculations \cite{chang2010}. Here, we mention the key features of ground-state auxiliary-field quantum Monte Carlo (AFQMC) methods that are relevant to this topic.

As the interaction in cold atoms is short-ranged compared to the interparticle spacing, the uniform Fermi gas can be modeled on a lattice by extended, attractive Hubbard model \cite{Hirsch1986}. The extended, attractive Hubbard model is written in second-quantized form as \cite{abe2004}

\begin{eqnarray}
  \hat{H}=\hat{K}+\hat{V} =  -t\overset{L}{\underset{\left\langle i,j\right\rangle \sigma}{\sum}}c_{i\sigma}^{\dagger}c_{j\sigma}+6t\overset{L}{\underset{i\sigma}{\sum}}c_{i\sigma}^{\dagger}c_{i\sigma}+U\overset{L}{\underset{i}{\sum}}n_{i\uparrow}n_{i\downarrow}, 
\label{eq:hamiltony}
\end{eqnarray}

Here $ L $ is the number of lattice sites, $ c_{i\sigma}^{\dagger} $ and $ c_{j\sigma} $ are creation and annihilation operators of an electron of spin $  \sigma $ on the ith lattice site, $ t=1 $ is the nearest-neighbor hopping energy, $ n_{i\sigma}=c_{i\sigma}^{\dagger}c_{i\sigma} $ is the density operator, and $ U $ is the on-site interaction strength. Two parameters, the interaction $ U/t $ and the particle density   $ \left(N_{\uparrow}+N_{\downarrow}\right) /L $, determine the physics given the topology of the lattice. The effective range expansion is
\begin{equation}
k\:cot\delta_{0}=-a^{-1}+\frac{1}{2}k^{2}r_{e}+..., 
\label{eq:ere}
\end{equation}
where $ a $ is the scattering length and $ r_{e} $ is the effective range. Since we are interested in the unitary limit, we adjust $ U $ to have $ a^{-1}=0 $. By solving the two-body problem of the model Eq.~(\ref{eq:hamiltony}) one finds that the scattering length diverges at $ U=-7.915\:t $ and $ r_{e}=-0.30572 $ \cite{werner}. We use these value of $ U $ and $ r_{e} $ throughout. The ground-state wave function $ \left|\psi_{0}\right\rangle $  can be obtained asymptotically from any trial wave function $ \left|\psi_{T}\right\rangle $  that is not orthogonal to $ \left|\psi_{0}\right\rangle $  by repeated applications of the ground-state projection 

\begin{equation}
\left|\psi_{0}\right\rangle \propto \lim_{\beta\rightarrow0}  exp\left(-\beta\left(\hat{H}-E_{T}\right)\right)\left|\psi_{T}\right\rangle,
\label{eq:gsp}
\end{equation}

Here $ E_{T} $ is guesses of the ground-state energy. The propagator may be evaluated using a Trotter-Suzuki approximation \cite{trotter1959,suzuki1976}
\begin{equation}
\left(e^{-\Delta\tau\left(\hat{K}+\hat{V}\right)}\right)^{n}=\left(e^{-\frac{1}{2}\Delta\tau\hat{K}}e^{-\Delta\tau\hat{V}}e^{-\frac{1}{2}\Delta\tau\hat{K}}\right)^{n}+\mathcal{O}\left(\Delta\tau^{2}\right),
\label{eq:trotter}
\end{equation}
Where $ \beta = \Delta\tau n $, and a Trotter error arises from the omission of the higher-order terms. The residual Trotter error can be controlled by extrapolation with several independent runs of sufficiently small $ \Delta\tau $ values, So, We will not be concerned by it here.

The Hubbard-Stratonovich (HS) transformation decoupled two-body propagator into one-body propagators by auxiliary fields \cite{hubbard1959,hirsch1983},
\begin{equation}
e^{-\Delta\tau\hat{V}}=\sum_{x}p\left(x\right)e^{\hat{O}\left(x\right)},
\label{eq:hs}
\end{equation}
where $ \hat{O}\left(x\right) $ is a one-body operator that depends on the auxiliary field $ x $ and $ p\left(x\right) $ is a probability density function with the normalization $ \sum p\left(x\right)=1 $. More precisely, the most commonly used HS transformation involves discrete auxiliary fields due to Hirsch \cite{hirsch1983}.
The spin form of this decomposition is 

\begin{eqnarray}
 	&&e^{-\Delta\tau Un_{i\uparrow}^{\dagger}n_{i\downarrow}}= \nonumber\\	&&e^{-\Delta\tau U\left(n_{i\uparrow}+n_{i\downarrow}-1\right)/2} \sum_{x_{i}
	\pm1}p\left(x_{i}\right)e^{\gamma x_{i}\left(n_{i\uparrow}+n_{i\downarrow}-1\right)},
\end{eqnarray}
Where $ cosh\left(\gamma\right)=exp\left(\Delta\tau\left|U\right|/2\right) $, which results in an Ising-like auxiliary field for each lattice site. We interpret $  p\left(x_{i}\right)=1/2 $ as a discrete probability density function (PDF) with $ x_{i}=\pm1 $. By setting

\begin{equation}
\hat{B}\left(x\right)=e^{-\frac{1}{2}\Delta\tau\hat{K}}e^{-\hat{O}\left(x\right)}e^{-\frac{1}{2}\Delta\tau\hat{K}},
\end{equation}

We can rewrite the projection as
\begin{equation}
\left|\psi_{0}\right\rangle =\sum_{\vec{X}}P\left(\vec{X}\right)\prod_{i=1}^{n}\hat{B}\left(x_{i}\right)\left|\psi_{T}\right\rangle .
\end{equation}
Where $ \vec{X} $ is $ \left(x_{1},x_{2},...,x_{n}\right) $, and $ P\left(\vec{X}\right)=\prod_{i}p\left(x_{i}\right) $. The ground-state energy can be obtained by
\begin{equation}
\left\langle \hat{A}\right\rangle _{0}=\frac{\left\langle \psi_{0}\left|\hat{A}\right|\psi_{0}\right\rangle }{\left\langle \psi_{0}|\psi_{0}\right\rangle },
\end{equation}

MC methods are used to calculate this many-dimensional integrals (3nL dimensions in the Hubbard model) by sampling the probability density function using the Metropolis algorithm \cite{metropplis1953} or a related method. 

The sign problem may occure when the integrand in the denominator, $ P\left(\vec{X}\right) \\ \left\langle \psi_{T}\left|\prod_{i=1}^{n}\hat{B}\left(x_{i}\right)\right|\psi_{T}\right\rangle $, is not positive. We represent the wave function at each stage by a finite ensemble of Slater determinants, i.e.
\begin{equation}
\left|\psi^{\left(0\right)}\right\rangle =\sum_{i}^{N_{w}}w_{i}^{\left(0\right)}\left|\phi^{\left(0\right)}\right\rangle ,
\label{eq:finite_ensemble}
\end{equation}
$ \left\{ \phi^{\left(n\right)}\right\} $  are Slater determinants will be sampled by a finite ensemble of points, $ N_{w} $. Each walker will have a weight $ w $ whose value is set as $ 1 $ at the beginning of projection.

In each step, we sample the auxiliary field $ x $ according to $ p\left(x\right) $ by MC and apply $ \hat{B}\left(x_{i}\right) $ to the Slater determinant wave function. Since the operators only contain one-body terms, they will generate another Slater determinant \cite{hamann1990}
\begin{equation}
\left|\psi^{\left(1\right)}\right\rangle =\sum_{i}^{N_{w}}\sum_{x_{i}}p\left(x_{i}\right)\hat{B}\left(x_{i}\right)w_{i}^{\left(0\right)}\left|\phi^{\left(0\right)}\right\rangle =\sum_{i}^{N_{w}}w_{i}^{\left(1\right)}\left|\phi^{\left(1\right)}\right\rangle.
\label{eq:evolution}
\end{equation}

As the random walk proceeds, some walkers may accumulate very large weights while some will have very small weights. By applying a population control bias \cite{zhang1997}, such that the overall probability distribution is preserved and the weights become more uniform. The different weights cause a loss of sampling efficiency by track of walkers that contribute little to the energy estimate. 

In Eq.~(\ref{eq:evolution}) multiplications of $ \hat{B}\left(x_{i}\right) $ to a Slater determinant lead to numerical instability, This instability is controlled by applying the modified Gram–Schmidt orthonormalization to each Slater determinant periodically \cite{zhang1997}.

In the standard ground-state AFQMC calculations, no information is contained in the sampling of $ \vec{X} $ on the importance of the resulting determinant in representing $ \left|\psi_{0}\right\rangle $. Computing the mixed estimator of the ground-state energy
\begin{equation}
E_{mixed}=\frac{\left\langle \phi_{T}\left|\hat{H}\right|\psi_{0}\right\rangle }{\left\langle \phi_{T}|\psi_{0}\right\rangle },
\label{eq:e_mixed}
\end{equation}
requires estimating the denominator by $ \sum_{i}\left\langle \phi_{T}|\phi_{k}\right\rangle $  where $ \left|\phi_{k}\right\rangle $  are random walkers after equilibration. Since walkers are sampled with no knowledge of $ \left\langle \phi_{T}|\phi_{k}\right\rangle $  , terms in the summation over $ \left|\phi_{k}\right\rangle $  can have large fluctuations that lead to large statistical errors in the MC estimate of the denominator, thereby in that of $ E_{mixed} $.

First, an importance function is defined, by importance sampling  \cite{huy2014}, 
\begin{equation}
O_{T}\left(\phi_{k}\right)\equiv\left\langle \phi_{T}|\phi_{k}\right\rangle ,
\label{eq:imp_sam}
\end{equation}
which estimates the overlap of a Slater determinant $ \left|\phi\right\rangle $  with the ground-state wave function (trial wave function). We then iterate an equivallent version of Eq.~(\ref{eq:evolution})

\begin{equation}
\left|\widetilde{\phi}^{\left(n+1\right)}\right\rangle =\sum_{x}\widetilde{P}\left(x\right)\hat{B}\left(x\right)\left|\widetilde{\phi}^{\left(n\right)}\right\rangle  ,
\label{eq:equi_ver}
\end{equation}

The walkers $ \left|\widetilde{\phi}^{\left(n\right)}\right\rangle $  are now sampled from a new distribution. They schematically represent the ground-state wave function by
\begin{equation}
\left|\psi^{\left(n\right)}\right\rangle =\sum_{i}^{N_{w}}w_{i}^{\left(n\right)}\frac{\left|\phi_{i}^{\left(n\right)}\right\rangle }{O_{T}\left(\phi_{k}^{\left(n\right)}\right)},
\end{equation}
same as Eq.~(\ref{eq:finite_ensemble}). The function $ \widetilde{P}\left(x\right) $ in Eq.~(\ref{eq:equi_ver})  is 
$ \widetilde{P}\left(x\right)=\prod_{i}^{M}\widetilde{p}\left(x_{i}\right) $, where the probability for sampling the auxiliary-field at each lattice site is given by
\begin{equation}
\widetilde{p}\left(x\right)=\frac{O_{T}\left(\phi_{k,i}^{\left(n\right)}\right)}{O_{T}\left(\phi_{k,i-1}^{\left(n\right)}\right)}p\left(x\right),
\end{equation}
Where $ \left|\phi_{k,i}^{\left(n\right)}\right\rangle =\hat{b}_{v}\left(x_{i}\right)\left|\phi_{k,i-1}^{\left(n\right)}\right\rangle  $ and $ \hat{b}_{v}^{\sigma}\left(x_{i}\right)=exp\left[-\left(\Delta\tau U/2-s\left(\sigma\right)\gamma x_{i}\right)c_{i\sigma}^{\dagger}c_{i\sigma}\right] $. Therefore, $ \widetilde{P}\left(x\right) $ is a function of current and future positions in Slater-determinant space. 

The sign problem occurs because of the fundamental symmetry between the fermion ground state $ \left|\psi^{\left(0\right)}\right\rangle $  and its negative $ -\left|\psi^{\left(0\right)}\right\rangle $ \cite{zhang1991}. This symmetry implies that, for any ensemble of Slater determinants $ \left\{ \left|\phi\right\rangle \right\} $  which gives a Monte Carlo representation of the ground-state wave function, there exists another ensemble $ \left\{ -\left|\phi\right\rangle \right\} $  which is also a correct representation. The projection would proceed identically if each random walker $ \left\{ \left|\phi\right\rangle \right\} $  were switched to $ \left\{ -\left|\phi\right\rangle \right\} $  at any given imaginary time, Thus the walker will make no further contribution to the representation of ground state because
\begin{equation}
\left\langle \psi_{0}|\phi\right\rangle =0\Rightarrow\left\langle \psi_{0}\left|e^{-\tau H}\right|\phi\right\rangle =0, 
\end{equation}
 For any $ \tau $. However, because the random walk has no knowledge of two divided degenerate halves of Slater determinant space, these paths continue to be sampled (randomly) in the random walk and become Monte Carlo noise.
 
 Constrained path approximation eliminates the decay of signal-to-noise ratio. It requires that each random walker at each step have a positive overlap with the trial wave function $ \left\langle \phi_{T}|\phi^{\left(n\right)}\right\rangle >0 $ and is easily implemented by redefining the importance function in  Eq.~(\ref{eq:imp_sam})
  \begin{equation}
 O_{T}\left(\phi_{k}\right)\equiv max\left\{ \left\langle \phi_{T}|\phi_{k}\right\rangle ,0\right\} .
  \end{equation}
  
 The ground-state energy for an ensemble $ \left\{ \left|\phi\right\rangle \right\} $  by the mixed estimator Eq.~(\ref{eq:e_mixed}), can be calculated by
 
 \begin{equation}
E_{mixed}=\frac{\sum_{k}w_{k}E_{L}\left[\phi_{T},\phi_{k}\right]}{\sum_{k}w_{k}},
 \end{equation}
  
 $ E_{L} $ is called local energy for any walker $ \phi $, and is not variational \cite{lieb1968},
 
 \begin{equation}
 E_{L}\left[\phi_{T},\phi\right]=\frac{\left\langle \phi_{T}\left|\hat{H}\right|\phi\right\rangle }{\left\langle \phi_{T}|\phi\right\rangle }.
 \end{equation}

\section{\label{result}Results}

We should point out that there are actually two different conventions for $ \xi_{N_{\uparrow} ,N_{\downarrow}} $ used in the literature. We refer to Eq.~(\ref{eq:en_ratio}) as the few-body definition for the energy ratio $ \xi_{N_{\uparrow},N_{\downarrow}} $ . This is the definition we use for all calculations presented here. The alternative definition for the energy ratio $ \xi_{N_{\uparrow},N_{\downarrow}} $ is thermodynamical definition. We can define the Fermi momenta and energies in terms of the particle density

\begin{equation}
k_{F,\uparrow\left(\downarrow\right)}=\left(6\pi^{2}\frac{N_{\uparrow\left(\downarrow\right)}}{L^{3}}\right)^{1/3}, E_{F,\uparrow\left(\downarrow\right)}=\frac{k_{F,\uparrow\left(\downarrow\right)}^{2}}{2m}.
\end{equation}

The ground-state energy of the noninteracting system at thermodynamic limit is 
$ \frac{3}{5}N_{\uparrow}E_{F,\uparrow}+\frac{3}{5}N_{\downarrow}E_{F,\downarrow} $. By using this, We define the thermodynamical energy ratio as below

\begin{equation}
\xi_{N_{\uparrow},N_{\downarrow}}^{thrmo}=\frac{E_{N_{\uparrow},N_{\downarrow}}^{0}}{\frac{3}{5}N_{\uparrow}E_{F,\uparrow}+\frac{3}{5}N_{\downarrow}E_{F,\downarrow}},
\end{equation}

$ CF_{N\uparrow N\downarrow} $ is the ratio between these two definitions, i.e.
\begin{equation}
CF_{N_{\uparrow}}=\frac{\xi_{N_{\uparrow},N_{\downarrow}}}{\xi_{N_{\uparrow},N_{\downarrow}}^{thrmo}}=\frac{9.1156\frac{N_{\uparrow}^{5/3}}{mL^{2}}}{E_{N_{\uparrow},N_{\downarrow}}^{0,free}}.
\end{equation}

The few-body ratio $ \xi_{N_{\uparrow},N_{\downarrow}} $ and thermodynamical ratio 
$ \xi_{N_{\uparrow},N_{\downarrow}}^{thrmo} $ differ due to shell effects in the noninteracting system. The values of $ CF_{N\uparrow N\downarrow} $ for several values of particle number with 
$ N\uparrow=N_{\downarrow} $ are tabulated in Table I in \cite{bour2011}.

\begin{figure}
\includegraphics[scale=0.72]{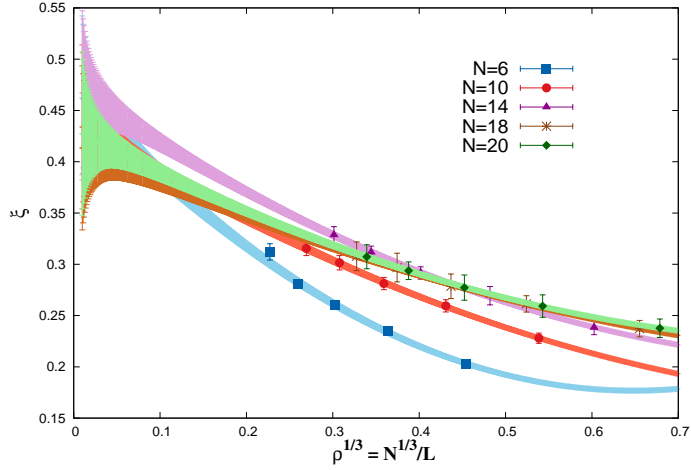} % Here is how to import EPS art
\caption{\label{fig:xilx} The calculated ground-state energy shown
as the value of $ \xi $ vs the lattice size for various particle numbers.}
\end{figure}

We have used \texttt{CPMC-Lab} package for numerical simulation \cite{huy2014}. In Fig.~\ref{fig:xilx}, we summarize our calculations of the energy as a function of $ \rho^{1/3} $ where $ \rho= {N}/{L^{3}} $, and the particle number is $ N=6,10,14,18, 20 $ and $L_{x}=L_{y}=L_{z}= 4,5,6,7,8 $. We plot $ \xi $, Eq.~(\ref{eq:en_ratio}), where we have in all cases used the infinite system free-gas energy 
$ E_{N\uparrow,N\downarrow}^{0,free} $. 

%PHYSICAL REVIEW A 83, 063619 (2011)

Our calculations show a significant size dependence. The fits are of the form \cite{carlson2011}, 
\begin{equation}
\frac{E_{N_{\uparrow},N_{\downarrow}}^{0}}{E_{N_{\uparrow},N_{\downarrow}}^{0,free}} = \xi_{0} + A \rho^{1/3} + B \rho^{2/3}.
\label{eq:xirho}
\end{equation}

The extrapolation in lattice size shows opposite slope as expected from opposite signs of their effective ranges. The values of $ \xi $ for extrapolation to $ \rho\rightarrow0 $, for $ N=6,10,14,18,20 $ are shown in Table~\ref{tab:extr_res}. Our error is a statistical standard errors.
%Our final error contains a statistical component and the errors associated with finite population sizes and finite time-step errors.
% This value is below previous experiments but is more compatible with recent experimental results of the Zwierlein group \cite{unpublished-full}.

 The behavior of energy as a function of $ k_{F}r_{e} $ for finite effective ranges is also examined. The slope of $ \xi $ is  universal in continuum Hamiltonians \cite{werner}
\begin{equation}
\xi\left(r_{e}\right) = \xi_{0} + S k_{F} r_{e}.
\label{eq:xikf}
\end{equation}
for the slope $ S $ of $ \xi $ with respect to finite $ r_{e} $. 

\begin{figure}
\includegraphics[scale=0.72]{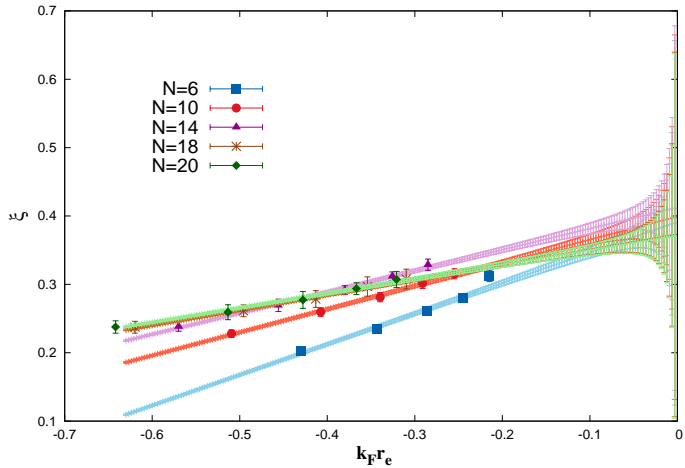} 
\caption{\label{fig:xikf} The ground-state energy as a function of $ k_{F}r_{e} $.}
\end{figure}

Fig.~\ref{fig:xikf} shows the AFQMC results for various values of the effective range. The values of $ \xi_{0} $ and $ S $ for extrapolation to $ k_{F}r_{e} \rightarrow 0 $ are given in  Table~\ref{tab:extr_res}. As we see from Table~\ref{tab:extr_res}, by increasing the number of particle, the decreasing of $ S $ becomes slower. Their trend are in agreement with results of $ S=0.12(0.01) $ for $ N = 66 $ \cite{carlson2011}.

From Table~\ref{tab:extr_res} we see that the energy ratio for the smallest system, $ N=6 $, is somewhat bigger than the rest. However the ratios for $ N\geq 10 $ are close to a central value of about $ 0.43 $.
Assuming no large changes to this ratio for $ N\geq 20 $, we estimate that $ \xi = 0.43(4) $.

\begin{table}[h]
\caption{\label{tab:extr_res}
The results for different  number of particles $ N $. $ \xi_{0}^{(1)} $ is the values of $ \xi $ for extrapolation to $ \rho \rightarrow 0 $ in Eq.~(\ref{eq:xirho}), and $ \xi_{0}^{(2)} $ is the values of $ \xi $ for extrapolation to $ k_{F}r_{e} \rightarrow 0 $ in Eq.~(\ref{eq:xikf}).   }
%\begin{ruledtabular}
\begin{tabular}{cccccc}
 $ N $            & $ N =6 $ & $ N =10 $ & $ N =14 $ & $ N =18 $    & $ N =20 $ \\
\hline
%$ \xi_{0}^{(1)} $ &0.4711996841   &0.4392401865  &0.4729065584  & 0.417423482  & 0.431819319  \\
$ \xi_{0}^{(1)} $ & 0.47(4) & 0.44(3)  & 0.47(3)  & 0.41(6)  & 0.43(5)  \\
%$ \xi_{0}^{(2)} $ & 0.3914088257  &0.3993141147  &0.4125307063  &0.3736459955  & 0.371342331  \\
$ \xi_{0}^{(2)} $ & 0.39(5) & 0.39(4)  & 0.41(4)  & 0.37(4)  & 0.37(5)  \\
$ S $             & 0.44(5) & 0.33(5)  & 0.30(4)  & 0.22(4)  & 0.21(5)  \\
\end{tabular}
%\end{ruledtabular}
\end{table}

 Our results, $ \xi = 0.43(4) $, might be useful as a comparison  with the fixed-node Green’s function Monte Carlo results
 $ \xi = 0.44(1) $ \cite{carlson2003}, $ \xi = 0.42(1) $ \cite{Astrakharchik2004} and almost with recent experimental results, $ 0.435(15) $ \cite{joseph2007}, $ 0.41(2) $ \cite{luo2009} and $ 0.41(1) $ \cite{navon2010} within its error. We have to point out that, because we consider small number of particles in our simulations, a careful extrapolation to the thermodynamic limit has to be done, otherwise, it is quite problematic and dangerous to make a direct comparison between the results and experimental measurements, as well as with other calculations. Using wave functions which restore symmetries of the system can reduce the systematic error significantly \cite{hao2013}.

It is quite important to point out that in Ref.\cite{carlson2011} similar calculations have been performed with a similar method. The main difference is that using the formalism of Ref. \cite{carlson2011} similar lattice calculations are sign problem free, and then the results are exact. 

\section{\label{summary}Summary}

We have tryed a lattice technique, constrained-path Monte Carlo method using the extended, attractive Hubbard model to treat strongly paired fermion systems. We have measured the ground-state energy for $ N $ 
spin $ 1/2 $- fermions in the unitary limit in a cubic. Our results at $ N=6,10,14,18,20 $ suggest that for large $ N $ the ratio of the ground-state energy to that of a free Fermi gas is $ 0.43(4) $.  

 Our simulation results might be useful as a comparison  with the fixed node Green's function Monte Carlo $ \xi = 0.44(1) $ \cite{carlson2003},
 $ 0.42(1) $ \cite{Astrakharchik2004}, novel lattice approach results $ 0.412(4) $ \cite{endres2013} and by a careful extrapolation to the thermodynamic limit can be compare  with recent experimental results, $ 0.435(15) $ \cite{joseph2007}, $ 0.41(2) $ \cite{luo2009} and $ 0.41(1) $ \cite{navon2010}. We also try to find results for universal dependence of the ground-state energy upon the effective range, $ S = 0.21(5) $ which can be compare with  $ S = 0.12(0.03) $ in Ref. \cite{carlson2011}. Ofcourse the
discrepancies of the results with others existing in literature should be deeply discussed somewhere. The method we describe should be useful with modification for the entire BCS-BEC transition \cite{Bagnato2015} and for probing many properties of cold Fermi gases. It can also be applied to atoms, molecules, solids, and correlated electron models, quantum chemistry and nuclear physics.

\begin{acknowledgement}
We are grateful for the authors and maintainers of \texttt{CPMC-Lab} \cite{huy2014}, a modified version of which is used for calculations in this work. We thank Professor Dean Lee and Dr. Hao Shi for useful discussions.
\end{acknowledgement}

\end{document}